\documentclass[12pt,subeqn,a4paper]{article}

\usepackage{amssymb,amsmath,amsfonts,amsthm, amscd, mathrsfs,helvet}
\usepackage{bm}
\usepackage{graphicx,verbatim}
\usepackage{psfrag}
\usepackage[all]{xy}
\usepackage{color}
\usepackage{epsfig}
\input epsf

\def\be{\begin{equation} }
\def\ee{\end{equation} }

\def\bea{\begin{eqnarray} }
\def\eea{\end{eqnarray}}

\newcommand{\f}[2]{\frac{#1}{#2}}
\newcommand{\ko}[1]{\left( #1 \right)}
\newcommand{\kko}[1]{\left[ #1 \right]}

\newcommand{\komoji}[1]{\mbox{$#1$}}

\def\eq{\equiv}
\def\be{\beta}

\def\ep{{\epsilon}}

\numberwithin{equation}{section}

\makeatletter

\newcommand{\lyxaddress}[1]{
  \par {\raggedright #1 
  \vspace{1.4em}
  \noindent\par}}

\makeatother

\begin{document}

\title{\begin{flushright}
{\small ITP -- Budapest Report 641}
\end{flushright}\vspace{1cm}
\textbf{\large Issues on magnon reflection}\large }

\author{L.~Palla\thanks{
palla@ludens.elte.hu
}, }

\maketitle

\lyxaddress{\centering \emph{Institute for Theoretical Physics
}\\
\emph{E\"{o}tv\"{o}s University} \emph{}\\
\emph{H-1117 Budapest, P\'{a}zm\'{a}ny P. s\'{e}t\'{a}ny 1 A, Hungary}
}

\begin{abstract}
Two questions related to reflections of magnons in AdS/CFT 
are discussed: namely the
problem of explaining the (physical) poles of the reflection amplitudes using
Landau type diagrams and the generalization of the Ghoshal-Zamolodchikov 
boundary state formalism to magnon reflections. 
\end{abstract}

\clearpage

\section{Introduction}

The recently discovered integrable structures \cite{MZ} in the planar ${\cal
  N}=4$ super Yang-Mills theory make possible to test the AdS/CFT
  correspondence \cite{Malda1} over the entire range of the 't Hooft
  coupling. The scaling dimensions of \lq infinitely long' 
single trace operators are computable to
all orders in perturbation theory by mapping the dilatation operator to an 
integrable spin chain \cite{BKS}. The spectrum is compared to the energy
  spectrum of a free closed string moving in ${\rm AdS}_5\times S^5$ with
  large angular momentum. In this limit the physical content of the theory is 
the spectrum of asymptotic states and their scattering matrix and they are
  highly constrained by the residual symmetries. Indeed Beisert showed
  \cite{beiserts} that the residual symmetries are sufficient to determine 
  the two particle scattering matrix between the elementary excitations (known
  as magnons) up to an overall scalar factor. This overall scalar factor,
  known as \lq\lq dressing factor'' plays an important role in interpolating
   the weak-strong coupling spectrum of the gauge/string correspondence
  \cite{AFS}, \cite{BHL}, \cite{Beisert:2006ez}. The dressing factor obeys an
  extra symmetry known as \lq\lq crossing'' imposed on the two particle
  scattering matrix \cite{BHL}, \cite{Beisert:2006ez}, \cite{Janik:2006dc}.       

 Recently Hofman and Maldacena \cite{HM} considered open strings attached to 
maximal giant gravitons \cite{Toumb} in ${\rm AdS}_5\times S^5$ (for related
  earlier work see \cite{BV}). They determined reflection amplitudes
  (sometimes also called boundary $S$ matrices) of the elementary magnons 
for two (integrability  preserving) cases, namely the $Y=0$ and the $Z=0$
  giant graviton branes.    
 For both cases 
the reflection amplitudes of the elementary magnons were determined
up to an overall phase factor in \cite{HM}
by exploiting the residual symmetries of the
problem. This phase factor for the $Y=0$ brane 
has recently been determined in \cite{ChenCor} from an analysis of the 
boundary crossing condition (BCC). The BCC for the $Z=0$ brane has also been 
determined in \cite{ChenCor} but the actual solution for the missing phase
factor of the elementary magnon reflection was given in \cite{ABR}. Using the
fusion method the authors of \cite{ABR} also determined the reflection
amplitudes for all the magnon bound states for both the $Y=0$ and the $Z=0$ 
branes. These results are in a certain sense summarized in \cite{AN1} where
the authors extend the Zamolodchikov-Faddeev algebra for open strings attached
to giant gravitons. 

In this paper two questions related to magnon reflections are investigated. 
The first concerns the interpretation of certain poles of the reflection
amplitudes in terms of the generalization of Landau diagrams. Some of the
first order poles (for the $Z=0$ brane) signal the presence of boundary bound
states \cite{HM}, \cite{ABR}, but not all the poles of the various reflection
amplitudes admit this interpretation 
and one of the aims of this paper is to suggest a possible explanation for
these \lq no boundary bound state' poles. This investigation is a
generalization to the boundary case of the   
recent program aimed   
to use the generalization of the (bulk) Coleman-Thun 
mechanism to explain the various (physical) poles in the scattering of 
 elementary magnons \cite{Dorey:2007xn} and magnon bound states \cite{DO}.

The second question we investigate is the generalization to magnon reflections
of the boundary state formalism originally worked out for relativistic 
boundary integrable theories in \cite{GhosZam}. 
The motivation for this investigation is at
least twofold: on the one hand this way a new derivation of the boundary
crossing condition (BCC) is obtained while on the other this formalism
naturally connects the magnon reflection
problem to the bulk mirror magnon theory introduced in \cite{AFmirror}.

The paper is organized as follows: in section 2 the basic facts about 
magnons are reviewed. In section 3 the poles in the magnon reflection
amplitudes on the $Y=0$ brane are exhibited. Section 4 is devoted to the
generalization of the boundary 
Coleman-Thun mechanism for the magnon problem; first 
some general remarks are made (valid also for magnon scattering in the bulk)
then the search for the appropriate Landau diagrams is described in details.
The generalization of boundary state formalism for magnon reflections is
described in section 5. We make our conclusions in sect. 6.

\section{Some basic facts about magnons}
 
\subsection{Kinematics and spectrum}

The fundamental excitations of the spin chain are the magnons, 
which form a sixteen-dimensional (short) BPS representation
of the unbroken $SU(2|2)\times SU(2|2)$ supersymmetry. The closure
of the SUSY algebra on this multiplet uniquely determines
\cite{beiserts} the magnon dispersion relation \cite{bds} (see also
\cite{Dispersion}),
\begin{equation}\label{disp}
E=\Delta-J=\sqrt{1+16g^{2}\sin^{2}\left(\frac{p}{2}\right)}
\end{equation}
where $g=\sqrt{g^{2}_{YM}N}/4\pi$. 
It is convenient to describe the magnons in terms of two complex spectral
parameters $x^\pm$. In terms of these parameters, 
the magnon momenta and energies
are expressed as, 
\begin{align}
p_{}&=p(x_{}^{\pm})=\f{1}{i}
\log\bigg(\f{x_{}^{+}}{x_{}^{-}}\bigg)\,,\label{p}\\
\ep_{}&=\ep(x_{}^{\pm})=\f{g}{i}\kko{\bigg(x_{}^{+}-\f{1}{x_{}^{+}}\bigg)-\bigg(x_{}^{-}-\f{1}{x_{}^{-}}\bigg)}\,.
\label{Delta}
\end{align}
The dispersion relation (\ref{disp}) is equivalent to the constraint
\begin{equation}
\bigg(x_{}^{+}+\frac{1}{x_{}^{+}}\bigg) - 
\bigg(x_{}^{-}+\frac{1}{x_{}^{-}}\bigg)=\frac{i}{g}\,.
\end{equation}

Furthermore, any number of
elementary magnons can form a stable bound state. The $Q$-magnon
bound state ($Q\in {\mathbb N}$) also belongs to a BPS representation of supersymmetry (of
dimension $16Q^{2}$). Therefore there is an infinite
tower of BPS states labeled by a positive integer $Q$ in the theory. The exact
dispersion relation for these states is again fixed by
supersymmetry to have the form \cite{doreymagone,doreymagfour},
\begin{equation}
E=\sqrt{Q^{2}+16g^{2}\sin^{2}\left(\frac{p}{2}\right)}
\label{dsp1}
\end{equation}
The spectral parameters of the constituent magnons in a $Q$-magnon
bound state are:    
\begin{equation}
x_{j}^{-} = x_{j+1}^{+}
\qquad \mbox{for}\quad 
j=\komoji{1,\dots, Q-1}\,.
\end{equation}
The resulting bound state is described by the spectral
parameters 
\begin{equation} 
X^{+}\equiv x_{1}^{+}\,,\quad 
X^{-}\equiv x_{Q}^{-}\,,
\end{equation}
and the total momentum $P$ and $U(1)$ charge $Q$ of the bound state are expressed as
\begin{align}
P(X^{\pm})&=\f{1}{i}\log\ko{\f{X^{+}}{X^{-}}}\,,\label{nagyp}\\
Q(X^{\pm})&=\f{g}{i}\kko{\ko{X^{+}+\f{1}{X^{+}}}-\ko{X^{-}+\f{1}{X^{-}}}}\label{Qkenyszer}\,.
\end{align}
One can also show the  energy
$E=\sum_{k=1}^{Q}\ep_{k}$ for the bound state is related to the 
spectral parameters $X^{\pm}$ through the expression
\begin{equation}
E(X^{\pm})      =\f{g}{i}\kko{\ko{X^{+}-\f{1}{X^{+}}}-\ko{X^{-}-\f{1}{X^{-}}}}\,,
        \label{E}
\end{equation}
while in terms of $P$ and $Q$\,,
\begin{equation}
E(P;Q)  =\sqrt{Q^{2}+16g^{2}\sin^{2}\ko{\f{P}{2}}}\,.\label{E2}
\end{equation}
The velocity of the particle in
appropriately-normalized world sheet coordinates
  $(x,t)$ is given as
\begin{equation}
v(X^\pm )=\frac{dx}{dt}=\frac{1}{2g}\frac{dE}{dP}=\frac{X^{+}+X^{-}}{1+X^{+}X^{-}}=
\frac{2g\sin(P)}{\sqrt{Q^{2}+16g^{2}\sin^{2}\left(\frac{P}{2}\right)}}.
\label{vel}
\end{equation}

\subsection{Bulk $S$ matrix}

In the infinite asymptotic spin chain limit the elementary magnons propagate
freely apart from pairwise scattering described by the two body scattering
matrix $S(x_1,x_2)$. It was shown by Beisert \cite{beiserts}, that by
demanding the invariance of $S(x_1,x_2)$ under the symmetry algebra it can be
constrained up to an overall scalar factor. Various versions of this $S$
matrix are in use in the literature depending on the choice of a basis. Here we
use the so called \lq\lq string'' basis \cite{AFZ} where   
\begin{equation}\label{ssmat}
S_{\rm full}=S_0^2(x_1,x_2)\bigl(\hat{S}_{su(2\vert 2)}(x_1,x_2)\otimes
\hat{S}_{su'(2\vert 2)}(x_1,x_2)\bigr).
\end{equation}
The flavour dependent $\hat{S}_{su(2\vert 2)}(x_1,x_2)$ and 
$\hat{S}_{su'(2\vert 2)}(x_1,x_2)$ 
parts are uniquely determined by the symmetry algebra and non trivially
satisfy the Yang-Baxter and unitarity equations. The scalar factor is given by
\cite{AFZ} 
\begin{equation}
S_0^2(x_1,x_2)=\frac{(x_1^--x_2^+)}{(x_1^+-x_2^-)}
\frac{(1-\frac{1}{x_1^+x_2^-})}
{(1-\frac{1}{x_1^-x_2^+})}{\sigma^2(x_1^\pm ,x_2^\pm )}
\equiv \frac{\tilde{S}(x_1^\pm ,x_2^\pm )}{A(x_1^\pm ,x_2^\pm )},
\quad A(x_1^\pm ,x_2^\pm )=\frac{(x_1^+-x_2^-)}{(x_1^--x_2^+)},
\end{equation}
where $\sigma^2(x_1^\pm ,x_2^\pm )$ is usually referred to as the \lq\lq
dressing factor''. The conjectured exact expression for this function \cite{Beisert:2006ez} is
conveniently given as an integral representation \cite{Dorey:2007xn},  
\begin{equation}
\sigma(x_{1}^{\pm}, x_{2}^{\pm})=\ko{\f{R(x_{1}^{+},
    x_{2}^{+})R(x_{1}^{-}, x_{2}^{-})}
{R(x_{1}^{+}, x_{2}^{-})R(x_{1}^{-}, x_{2}^{+})}}\,,\qquad 
R(x_{1}, x_{2})=e^{i\kko{\chi(x_{1}, x_{2})-\chi(x_{2}, x_{1})}}\,,
\label{BEHLS}
\end{equation}
where 
\begin{equation}
\chi(x_{1},x_{2})=-i\oint_{\mathcal C}\f{dz_{1}}{2\pi}\oint_{\mathcal
  C}\f{dz_{2}}{2\pi}
\f{\log\Gamma\ko{1+ig\ko{z_{1}+\f{1}{z_{1}}-z_{2}-\f{1}{z_{2}}}}}
{(z_{1}-x_{1})(z_{2}-x_{2})}\,,
\label{chi}
\end{equation}
with the contours in (\ref{chi}) being unit circles $|z_{1}|=|z_{2}|=1$.

Now consider the highest weight ($(1,1)$) component of the magnon $\phi
(x_1)$,  
that scatters diagonally, i.e. for which the full scattering amplitude can be 
written as
\begin{equation}\label{egymgsz}
\vert\phi (x_1)\phi (x_2)\rangle = A(x_1^\pm ,x_2^\pm ) 
\tilde{S}(x_1^\pm ,x_2^\pm )\vert\phi (x_2)\phi (x_1)\rangle .
\end{equation}
(Note that - apart from the $\sigma^2(x_1^\pm ,x_2^\pm )$ factor - 
the product $A(x_1^\pm ,x_2^\pm )\tilde{S}(x_1^\pm ,x_2^\pm )$ is 
nothing but the BDS piece of the S-matrix 
$S_{\rm BDS}(x_1^\pm ,x_2^\pm )$). Then, because of factorization of the multiparticle
S-matrix, the two body scattering matrix between magnon bound states
$\Psi_{Q_1}(X_1)$, $\Psi_{Q_2}(X_2)$ with spectral parameters
$X_1^\pm$ and $X_2^\pm$,  
consisting of $Q_1$ and $Q_2$  pieces of $\phi$  
respectively ($Q_1\geq Q_2$), is 
simply obtained as the product of two body S-matrices describing all possible
pair-wise scattering between the constituent magnons (\lq\lq fusion
procedure''). The outcome is
\begin{equation}\label{multimgsz}
 \vert\Psi_{Q_1} (X_1)\Psi_{Q_2} (X_2)\rangle 
= A(X_1^\pm ,X_2^\pm ) 
\tilde{S}(X_1^\pm ,X_2^\pm )\prod\limits_{k=0}^{Q_2-1}F(X_1^\pm ,X_2^\pm ,k)
\vert\Psi_{Q_2} (X_2)\Psi_{Q_1} (X_1)\rangle ,
\end{equation}
with
\begin{equation}
F(X_1^\pm,X_2^\pm,k)=\left(\frac{X_1^++\frac{1}{X_1^+}-X_2^+-\frac{1}{X_2^+}+\frac{ik}{g}}
{X_1^-+\frac{1}{X_1^-}-X_2^--\frac{1}{X_2^-}-\frac{ik}{g}}\right)^{2-\delta_{k,0}}.
\end{equation}
(This form of $F(X_1^\pm,X_2^\pm,0)$ is valid for $Q_1>Q_2$; if $Q_1=Q_2$ 
then $F(X_1^\pm,X_2^\pm,0)\equiv 1$).
The scattering phases appearing in eq.(\ref{egymgsz}, \ref{multimgsz}) have
first and second order poles. Some of the poles (like the first order ones 
at $X_1^-=X_2^+$) signal the possibility of forming bound states but not all 
of them can be explained this way 
and a program to explain them in terms of the particles in the spectrum using
the generalization of Landau diagrams was initiated in \cite{Dorey:2007xn} 
\cite{DO}. 

\subsection{Physicality conditions}

In general the singularities of the bulk $S$ matrix or the reflection
amplitudes occur at complex values of the external momenta and energies. Of
these singularities only the ones in the \lq physical domain' require an
explanation in terms of the particles (and boundary bound states) in the
spectrum. In ref.\cite{Dorey:2007xn}\cite{DO} the following condition was
proposed to decide the \lq physicality' of a bulk 
singularity: it is physical if parametrically 
it comes close to the positive real energy axis in any of the following three
limits: 
\begin{description}
\item[{(i)}] {The Giant Magnon limit}\,:~ $g\to \infty$ while $P$ kept fixed, 
when 
\begin{equation}
X^{+}\simeq 1/X^{-}\simeq e^{iP/2}\,,\quad 
\quad 
E\simeq 4g\sin\ko{\f{\vert P\vert}{2}}\,.
\end{equation}
\item[{(ii)}] {Plane-Wave limit}\,:~ $g\to \infty$ with $k\eq 2gP$ kept fixed, 
when  
\begin{equation}
X^{+}\simeq X^{-}\simeq \f{Q+\sqrt{Q^{2}+k^{2}}}{k}\in {\mathbb R}\,,\quad 
\quad 
E\simeq \sqrt{Q^{2}+k^{2}}\,.
\label{pp-wave region}
\end{equation}
\item[{(iii)}] {Heisenberg spin-chain limit}\,:~ $g\ll 1$ limit, when 
\begin{equation}
X^{\pm}\mp \f{iQ}{2g}\simeq \f{1}{2g}\cot\ko{\f{P}{2}}\,,\quad 
E\simeq Q+\f{8g^{2}}{Q}\sin^{2}\ko{\f{P}{2}}\,.
\end{equation}
\end{description}
In the following we accept this condition also for singularities in the
reflection amplitudes.

\section{(Multi)magnon reflections}

In integrable field theories new phenomena appear when they are restricted to a
half line with some non trivial boundaries, the boundary sine-Gordon model 
being a well studied prime example \cite{GhosZam}. 
Usually the integrability of the
bulk theory is preserved for some special boundary conditions only. A similar 
thing happens in the AdS/CFT correspondence, where on the string theory
side D-branes introduce non trivial boundaries of the string world sheet,
while in the ${\cal N}=4$ super Yang-Mills side (sub)determinant fields
introduce boundaries to composite operators. Recently, in \cite{HM}, two 
special (integrable) boundary conditions of the open spin chains in 
${\cal N}=4$ super Yang-Mills were investigated which describe giant
gravitons interacting with the elementary magnons of the chain attached to
it. The first case,  the $Y=0$ brane is represented by composite operators
containing a determinant factor det($Y$), while the second, the $Z=0$ brane is 
represented by composite operators containing det($Z$). ($W,Y,Z$ denote the
three complex scalar fields of the ${\cal N}=4$ super Yang-Mills). The
essential difference between the two cases is that in the  $Z=0$ brane the
open super Yang-Mills spin chain is attached to the giant graviton through
some 
boundary impurities $\chi $, $\chi ``$, and as a result it has a boundary
degree of freedom, while the $Y=0$ brane has no boundary degrees of freedom.

\subsection{Reflections on the $Y=0$ brane}

We start by reviewing the (multi)magnon reflection amplitudes on the $Y=0$
brane. The reflection amplitude for the elementary magnon component $\phi$
(which is a singlet under the residual $su(1\vert 2)\otimes su(1\vert 2)$
symmetry) is given by 
\begin{equation}\label{egymagrefl}
R_{R\ {\rm full}}^Y:\quad \vert\phi (x^\pm)\rangle =
-\sigma (x^\pm ,-x^\mp )\vert\phi (-x^\mp)\rangle ,\quad 
R_{L\ {\rm full}}^Y : R_{R\ {\rm full}}^Y(x^\pm\rightarrow -x^\mp )         
\end{equation}
for reflections on the left (respectively right) boundaries. Since they are
obtained from each other by the parity transformation ($x^\pm\rightarrow -x^\mp
$) in the following we concentrate on reflections from the right boundary
only. For a magnon bound state $\Psi_{Q}(X)$ the fusion procedure 
(that now involves also the reflection of elementary magnons 
(\ref{egymagrefl})) gives the
total reflection amplitude $R_{R\ {\rm full}}^Y$ as
\begin{equation}
\vert\Psi_{Q} (X^\pm)\rangle =
R_{Q\,R}(X)\vert\Psi_{Q} (-X^\mp)\rangle ,
\end{equation}
where
\begin{equation}\label{multimagrefl} 
R_{Q\,R}(X)=- \sigma^{-1}(-X^\mp ,X^\pm )\prod\limits_{k=1}^{Q-1}
\left(\frac{X^++\frac{1}{X^+}-\frac{ik}{2g}}{X^-+\frac{1}{X^-}+\frac{ik}{2g}}\right)\,.
\end{equation}
Using the representation (\ref{BEHLS},\ref{chi}) one can show that
$R_{Q\,R}(X)$ has only first order poles and zeroes given by 
the following expressions 
\begin{align}
{\rm poles}\qquad {\rm at}\qquad X^-+\frac{1}{X^-}&=i\frac{m}{2g}\qquad
m=-(Q-1),\dots ,-1,1,2,\dots \label{polus}\\ 
{\rm zeroes}\qquad {\rm at}\qquad X^++\frac{1}{X^+}&=i\frac{l}{2g}\qquad
l=(Q-1),\dots ,1,-1,-2,\dots \,,\label{nulla}
\end{align}
where, for both poles and zeroes, the second (infinitely long) set originates 
from the dressing factor. It was remarked in \cite{ABR} that the first order
poles in (\ref{polus}) can not be identified with the formation of boundary
bound states as they would not solve the boundary Bethe-Yang equation. Part
of the aims of this paper is to outline an alternative explanation of these
first order poles.
 
To check the \lq\lq physicality'' of the poles in (\ref{polus}) we determine 
$X^\pm$ for the poles by combining (\ref{Qkenyszer}) and (\ref{polus}):
\begin{align}
X^+&=\frac{i}{4g}\left( 2Q+m+\sqrt{16g^2+(2Q+m)^2}\right) \label{yp1} \\
X^-&=-\frac{i}{4g}\left( \sqrt{16g^2+m^2}-m\right)\label{yp2}\;.
\end{align}
Using them in (\ref{E}) gives the energy of the $Q$ magnon bound state at the
pole as
\begin{equation}
E=\frac{1}{2}\left(\sqrt{16g^2+(2Q+m)^2}+\sqrt{16g^2+m^2}\right)\longrightarrow
4g\qquad {\rm for}\quad g\rightarrow\infty ,
\end{equation}
while the momentum of the bound state (\ref{nagyp}), 
can conveniently be parameterized as
$P=\pm\pi-iq$ with
\begin{equation}
q=\ln \frac{2Q+m+\sqrt{16g^2+(2Q+m)^2}}{\sqrt{16g^2+m^2}-m}\sim \frac{Q+m}{2g}
\qquad {\rm for}\quad g\rightarrow\infty .
\end{equation}
 These expressions show that in the giant magnon regime the pole satisfies the
 physicality condition. For later reference we also note that the velocity of
 the bound state at the pole can be written as
\begin{equation}\label{polusseb}
v(X)=\frac{1+\frac{X^+}{X^-}}{X^++\frac{1}{X^-}}=-\frac{4g}{i}\frac{L}{N},
\qquad L>0,\ N>0,
\end{equation}
indicating that it is indeed heading towards the right boundary.

\section{Boundary Coleman-Thun mechanism for magnons}

Explaining higher order poles in the exact S-matrices of integrable 1+1
dimensional models goes back to the work of Coleman and Thun \cite{ColThu},
who were the 
first to realize that in 1+1 dimensions some anomalous thresholds may appear
in the form of these poles. An interesting aspect of this analysis was the
realization, that sometimes even first order poles may be explained as
anomalous thresholds, since by this they broke the usual association of 
first order poles in the S-matrix with a particle state in either the forward
or the crossed channel. In ref. \cite{DTW} \cite{MD} and \cite{BPTT} it was
found that this Coleman-Thun mechanism  works also in the presence of
(integrability preserving) boundaries, as several first order poles of the
various reflection amplitudes - instead of describing boundary bound states -
could be explained in terms of on-shell (anomalous threshold) diagrams for
multiple scattering processes now involving also the reflections on the
boundary\footnote{The complete generalization of the underlying Landau
  equations for any (not necessarily integrable) relativistic boundary theory 
can be found in \cite{BBT1} \cite{BBT2}}. 
It is important to emphasize that normally an anomalous threshold
diagram would lead to a pole of order higher than one (for a diagram with $N$
internal lines and $L$ loops the order is $N-2L$) 
and a first order pole
is obtained either after taking into account the combination of several
diagrams or because one or more \lq\lq internal'' 
reflection/scattering amplitudes develop zeroes exactly when the diagram goes
on-shell. 

 One
of the 
aims of this paper is to generalize the boundary Coleman-Thun mechanism for
(multi)magnon reflections.   

\subsection{General remarks on Coleman-Thun mechanism for magnons}

The Coleman-Thun mechanism is quantum field theoretic in nature as it relys on
the Landau equations that determine the (necessary) conditions for a Feynman
diagram to develop a singularity. To derive these equations \cite{IZ} 
one assumes the
usual (Minkowski space) form of the internal propagators in addition to
conservation of energy and momentum at the internal vertices. Assuming that
there is a (obviously non relativistic) field theory underlying the
(multi)magnon scattering/reflections one may try to generalize the Landau
equations for the magnon problem. In doing so one has to use the explicit 
form of the propagator on the internal lines. A natural choice is to 
take it in the form   
\begin{equation}\label{magnonprop}
\Pi (E,P)=\frac{i}{E^2-16g^2\sin^2\frac{P}{2}-Q^2+i\epsilon}
\end{equation}
for a magnon(bound state) with energy $E$ and momentum $P$,  
since this is in accord with 
the dispersion relations (\ref{Delta},\ref{E2}), and its denominator 
is also quadratic in the
energy as in the relativistic case. Furthermore, after appropriate analytical 
continuations, this form may also describe the free propagator of the mirror 
magnon model \cite{AFmirror}.  
Since energy and
momentum are conserved at the vertices also for the magnon problem,  
using also (\ref{magnonprop}) 
one can repeat the procedure in \cite{IZ} with the outcome that 
the Landau
equations for any diagram with $I$ internal lines and $L$ loops
take the form\footnote{We consider the leading singularity of the diagram only}:
\begin{equation}\label{Landau1}
E_j^2-16g^2\sin^2\frac{P_j}{2}-Q_j^2=0,\qquad j=1,\dots I, 
\end{equation}
for all the internal lines, and
\begin{equation}\label{Landau2}
\sum\limits_{i\in L_l}\alpha_iE_i=0,\qquad 
-8g^2\sum\limits_{i\in L_l}\alpha_i\sin P_i=0,,\qquad l=1,\dots L
\end{equation}
for all the loops $L_l$. Eq.(\ref{Landau1}) means of course that for the
singularity all the internal lines must go on-shell, but eq.(\ref{Landau2}) 
have more interesting consequences if we want the singularities to correspond
to spacetime diagrams with the vertices (representing local
interaction regions) being points in spacetime. Indeed in view of 
eq.(\ref{Landau2}) the diagram representing
the singularity becomes closed if the internal 
line connecting two vertices is
determined as 
\begin{equation}\label{vonal}
x_1^0-x_2^0=\alpha E,\qquad x_1^1-x_2^1=\alpha \sin P\,,
\end{equation}          
instead of the usual expression \cite{IZ}, where $P$ would appear instead of
$\sin P$ on the r.h.s of the second 
equation in (\ref{vonal}).\footnote{This subtle
difference seems to have been unnoticed in the earlier works
\cite{Dorey:2007xn} \cite{DO}.} Therefore the 
Landau diagrams in the magnon problem (i.e. spacetime diagrams representing
the singularity) 
may have the same topology as the
ordinary Landau diagrams but - unlike in the ordinary case - 
they cannot be interpreted as the propagation of
on shell particles. (Note in particular, that the two metrics corresponding 
to the dispersion relations (\ref{Delta},\ref{E2}) on the one hand 
and the one on the lines (\ref{vonal}) on the other 
are different while in the ordinary case they are the same).\footnote{It is 
interesting to note how the Landau equations change if one assumes the non 
relativistic propagator
$$
G(E,P)=\frac{i}{E-\sqrt{Q^2+16g^2\sin^2\frac{P}{2}}+i\epsilon}
$$
instead of $\Pi(E,P)$ for (multi)magnons. In this case one finds the loop 
equations    
$$
\sum\limits_{i\in L_l}\alpha_i=0,\qquad 
-4g\sum\limits_{i\in L_l}\alpha_iv(P_i)=0,,\qquad l=1,\dots L
$$
where $v(P_i)$ is the particle's velocity, eq.(\ref{vel}). These equations
admit no obvious space time interpretation in spite of the clear physical
meaning of the second set.}

In the presence of reflecting boundaries eq.(\ref{Landau1},\ref{Landau2})
remain valid and the only extra restriction is that at the vertex describing
the reflection only the particle's energy is conserved while its momentum
changes sign, i.e. at the reflection vertex the spectral parameter of the
particle changes as $X^\pm\rightarrow -X^\mp$. 

\subsection{Landau diagrams for reflections on the $Y=0$ brane} 

In the following we look for (boundary) Landau diagrams that could explain the 
first order poles listed in (\ref{polus}). To start with we give here a sample
of $0$, $1$ and $2$ loop diagrams used earlier to describe several poles in
various reflection amplitudes of the boundary sine-Gordon model 
\cite{MD} \cite{BPTT}:
  
\vspace{.5cm}~~\includegraphics[%
  width=7cm,
  height=5cm]{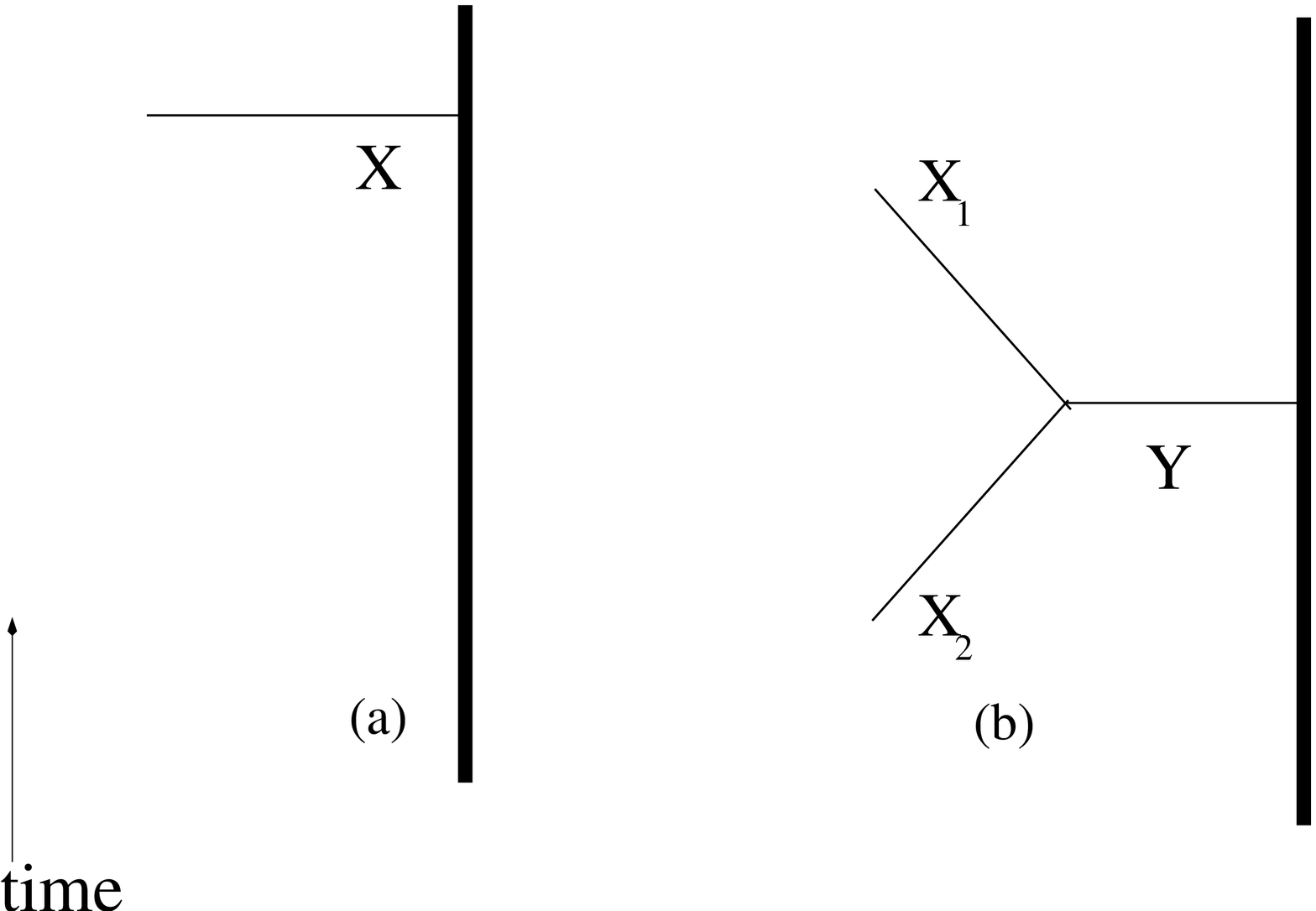}~~~~~~~~~~~\includegraphics[%
  width=7cm,
  height=5cm]{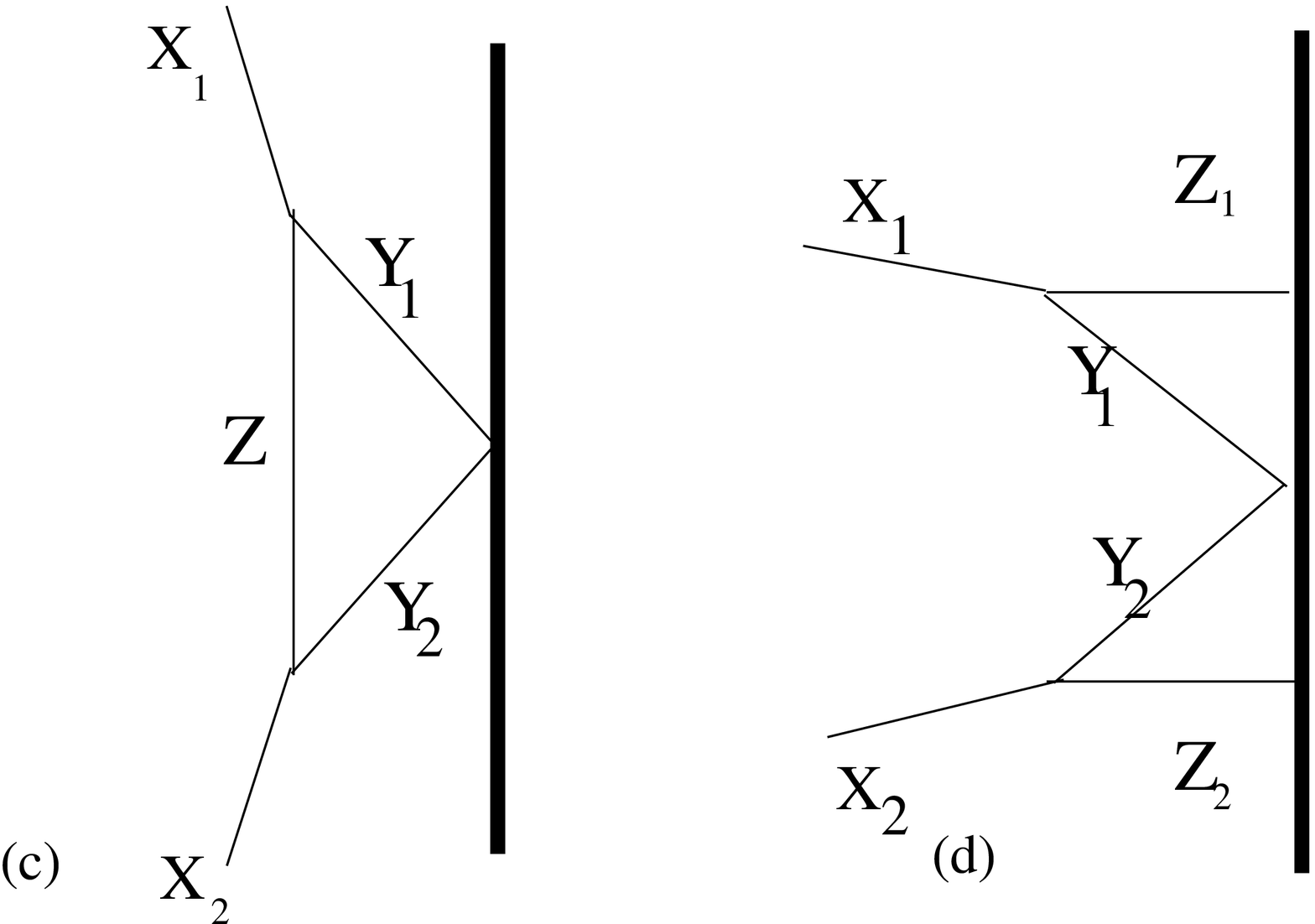}
  
Here we adapted the diagrams to the magnon problem by characterizing the 
magnon(bound states) with their spectral parameters. At the bulk vertices 
energy, momentum and $U(1)$ charge are conserved and these are so restrictive, 
that the ($X,Y$ and $Z$) spectral parameters of the three particles joined by
the vertex should be related to each other by one of the following 
\lq vertex conditions' \cite{Dorey:2007xn}\cite{DO} 
\begin{align}
X^-&=Z^-,\qquad X^+=Y^+,\qquad Z^+=Y^-\\
X^+&=Z^+,\qquad X^-=Y^-,\qquad Z^-=Y^+\\
X^-&=Z^-,\qquad X^+=\frac{1}{Y^-},\qquad Z^+=\frac{1}{Y^+}\\
X^+&=Z^+,\qquad X^-=\frac{1}{Y^+},\qquad Z^-=\frac{1}{Y^-}
\end{align}
The question which of these and in what situation may correspond to bound
state poles of the bulk $S$ matrix is discussed at length in 
\cite{Dorey:2007xn}\cite{DO}.
Because of the various possibilities to every diagram one should also tell the
vertex conditions chosen at the bulk vertices. 
Therefore the following algorithm is devised to check all the candidate diagrams:
\begin{enumerate}
\item choose one of the admissible vertex conditions at every bulk vertex,
\item impose the reflection condition at the vertex on the boundary, 
\item check that as a result one gets indeed the reflection of the external
  legs $X_1^\mp =-X_2^\pm$, 
\item check whether $Q$ for the internal lines - as determined from
  (\ref{Qkenyszer}) - is physical ($Q\geq 0$),
\item check that at all (bulk) vertices the particles are indeed heading
  towards/away from the boundary by comparing their velocities to 
(\ref{polusseb}),
\item check whether the internal reflection/scattering amplitudes have zeroes
  or poles to modify the naive counting of the degree of the pole ($N-2L$).
\end{enumerate} 
In checking the last three points it is assumed of course that we are at
the pole in question, i.e. the incoming $X_2^\pm $ is 
  determined by the location of the pole (\ref{polus}) and by
  (\ref{Qkenyszer}).  

On several diagrams of the figure there are horizontal lines, corresponding to
$E=0$ magnon(bound states) according to (\ref{vonal}). The spectral parameters 
of such a particle satisfy
\begin{equation}
(X^+-X^-)(1+\frac{1}{X^+X^-})=0.
\end{equation}
The $X^+=X^-$ solution gives vanishing $Q=0$, therefore we must choose the
other, and this, when combined with (\ref{Qkenyszer}) gives
\begin{equation}
X^-+\frac{1}{X^-}=-\frac{i}{2g}Q,\quad\qquad 
X^++\frac{1}{X^+}=\frac{i}{2g}Q
\end{equation}
The fact that this $X^-$ is different from the ones in (\ref{polus})
indicates that no poles may correspond to a particle \lq\lq standing
perpendicular'' to the boundary (i.e. diagram (a) is ruled out). This may be
understood in the following heuristic way: recently, in ordinary boundary
integrable QFTs, it was shown, that the existence of a pole corresponding to a
particle standing perpendicular to the boundary is related to some fields
developing non trivial vacuum expectation values (vev)
\cite{BPT2}. Assuming this relation
exists also in the field theory underlying the 
magnon problem the absence of this pole is in accord with
the residual supersymmetry that rules out non trivial vev-s.    

Diagram (b) is also ruled out: using any of the vertex conditions and
insisting on the reflection of the external leg $X_1^\mp=-X_2^\pm$ gives a
system of equations that can simultaneously be satisfied only if $X_2^\pm$
satisfy additional requirements (like $X_2^+=-X_2^-$); which the actual
solutions (\ref{yp1},\ref{yp2}) 
fail to fulfill. A similar - but more involved - argument rules out
diagram (d).  

The triangle diagram (diagram (c)), providing naively a first order pole,  
is interesting as it contains a \lq\lq
vertical'' line i.e. one which runs parallel to the boundary. The physically
interesting value for the momentum of the corresponding particle is
$P=\pm\pi$\footnote{The $P=0$ other solution leads to $Z^+=Z^-$ and $Q=0$,
  thus we discard it.}, and this gives for the spectral parameters
$Z^+=-Z^-$. (It is shown in the following, that the actual solutions of the
first three steps of the algorithm outlined earlier indeed contain such a
line). Then it is interesting to note that eq.(\ref{vonal}) can obviously be
satisfied since $Y_1^\pm=-Y_2^\mp$ implies $P_1^Y=-P_2^Y$ and $E_1^Y=E_2^Y$, 
thus choosing $\alpha_1=\alpha_2=\alpha $ solves the second equation in 
(\ref{vonal}) for any value of $\alpha$, while the first requires only 
$2\alpha E_1^Y+\alpha_3 E^Z=0$.  

The outcome of the study of the various triangle diagrams is summarized in
Table 1. Here we collected only those that after imposing the reflection at 
the vertex on the boundary ($Y_1^\pm=-Y_2^\mp$) 
 lead to the reflection of the external legs $X_1^\pm=-X_2^\mp$. For all of
 them one also gets $Z^+=-Z^-$ as promised earlier. In each case the vertex
 conditions used at the upper (lower) vertices on diagram (c) 
are listed first (second). The second column contains the $Q$ values for the
internal lines determined from the conservation equations and from $X_2^\pm$ 
being determined by the pole condition (\ref{polus}) and (\ref{Qkenyszer}):
\begin{equation}
X_2^-+\frac{1}{X_2^-}=i\frac{m}{2g},\qquad
X_2^++\frac{1}{X_2^+}=i\frac{2Q+m}{2g},\qquad
m=-(Q-1),\dots ,-1,1,2,\dots
\end{equation}
The second and fourth possibilities are ruled out since for the $Y$ lines 
here we
find the unphysical value  $Q_{Y_1}=Q_{Y_2}\equiv Q_Y<0$. The third
possibility is acceptable for negative $m$-s: $m\in -(Q-1),\dots -1$. In the
third column the $Y_2^\pm$ spectral parameters of the internal particle
reflecting on the boundary are collected. These are useful when checking
whether the internal reflection amplitude has a zero or a pole which would
modify the naive degree of the diagram. To do this one has to compare these
values to the ones in (\ref{polus}) and (\ref{nulla}) where the substitution 
$Q\rightarrow Q_Y$ is made. From this it turns out that for the first
possibility the internal reflection amplitude has a zero at this particular
values of the spectral parameters thus rendering the diagram {\sl finite},
while for third possibility there is a pole so that the diagram gives finally
a {\sl second order pole}. Thus the first diagram can not be used to explain
the poles in (\ref{polus}) while the third may perhaps be combined with other
diagrams giving also second order poles.    
 
\begin{table}
\caption{\small The four triangle diagrams leading to the reflection of the
  external legs.}
\begin{center}
\begin{tabular}{|c|c|c|}\hline
vertex conditions & $Q$ values & reflecting magnon\\ \hline
$X_1^-=Z^-$,\,$X_1^+=\frac{1}{Y_1^-}$,\,$Z^+=\frac{1}{Y_1^+}$ & $Q_Z=2Q+m$ & 
$Y_2^++\frac{1}{Y_2^+}=\frac{im}{2g}$\\
$X_2^+=Z^+$,\,$X_2^-=\frac{1}{Y_2^+}$,\,$Z^-=\frac{1}{Y_2^-}$ & $Q_Y=Q+m$ & 
$Y_2^-+\frac{1}{Y_2^-}=-\frac{i}{2g}(2Q+m)$\\ \hline
$X_1^+=Z^+$,\,$X_1^-=\frac{1}{Y_1^+}$,\,$Z^-=\frac{1}{Y_1^-}$ & $Q_Y=-(Q+m)$ & 
 \\
$X_2^-=Z^-$,\,$X_2^+=\frac{1}{Y_2^-}$,\,$Z^+=\frac{1}{Y_2^+}$ &  & 
\\ \hline
$X_1^+=Z^+$,\,$X_1^-={Y_1^-}$,\,$Z^-={Y_1^+}$ & $Q_Z=-m$ & 
$Y_2^++\frac{1}{Y_2^+}=\frac{i}{2g}(2Q+m)$\\
$X_2^-=Z^-$,\,$X_2^+={Y_2^+}$,\,$Z^+={Y_2^-}$ & $Q_Y=Q+m$ & 
$Y_2^-+\frac{1}{Y_2^-}=-\frac{im}{2g}$\\ \hline
$X_1^-=Z^-$,\,$X_1^+={Y_1^+}$,\,$Z^+={Y_1^-}$ & $Q_Y=-(Q+m)$ & 
 \\
$X_2^+=Z^+$,\,$X_2^-={Y_2^-}$,\,$Z^-={Y_2^+}$ &  & 
\\ \hline
\end{tabular}
\end{center}
\end{table}

Finally, to complete the study of the triangle diagrams we give here the
velocity of the particle $Y_2^\pm$ at the lower vertex on diagram (c) for the
first and third cases\footnote{the velocity of the particle $Y_1$ at the upper vertex
is automatically opposite to this one}:
\begin{equation}\label{ketseb}
{\rm first:}\quad v(Y_2)=\frac{1-\frac{X_2^+}{X_2^-}}{\frac{1}{X_2^-}-X_2^+}, 
\qquad\qquad  
{\rm third:}\quad v(Y_2)=\frac{1-\frac{X_2^+}{X_2^-}}{X_2^+-\frac{1}{X_2^-}}.
\end{equation}
Obviously only one of them can have the same sign as $v(X_2)$,
(\ref{polusseb}), and using the actual values of $X_2^\pm$ in
(\ref{yp1},\ref{yp2}) reveals that it is the {\sl first} diagram where this
happens. Therefore the third diagram is also ruled out and we conclude that
there is no simple triangle diagram that could be used to explain the first
order poles (\ref{polus}). 

The way out is to consider appropriate two loop diagrams obtained by combining 
in a certain sense the simple triangles of Table 1 where the zero of the
reflection amplitude renders the diagram first order. The diagram that works

\vspace{.5cm}~~~~~~~~~~~~~~~~~~~~~~~~~~~~~~~~~~~~~~~~~~~~~\includegraphics[%
  width=3.5cm,
  height=5cm]{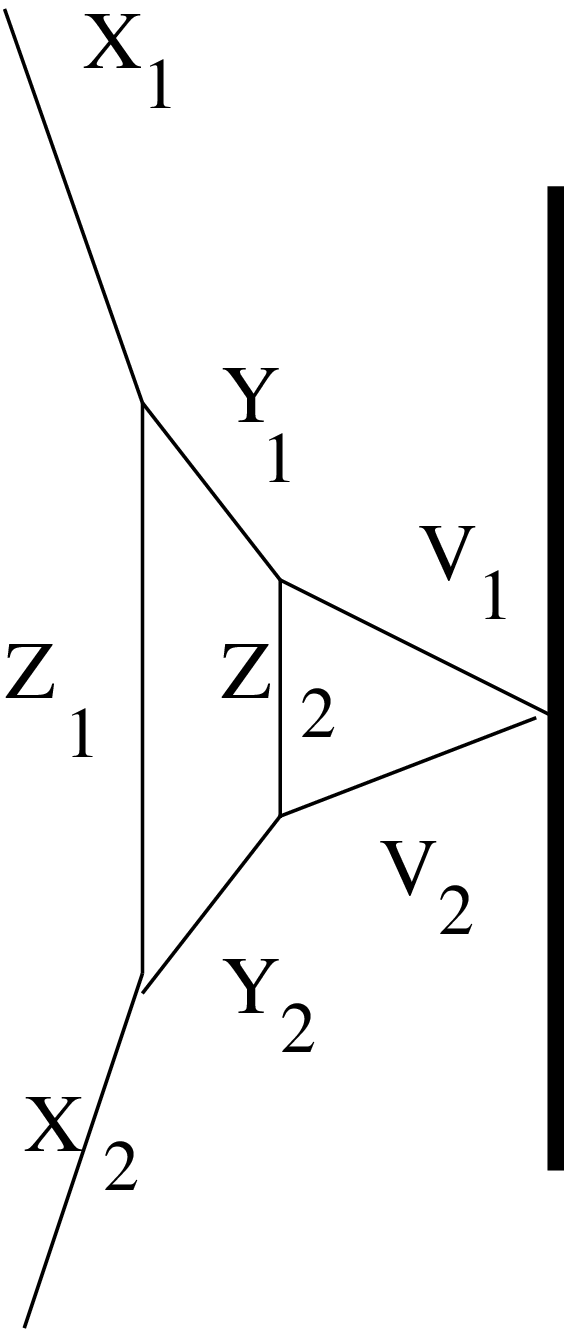}

\noindent has the same upper and lower 
vertex conditions along the longer vertical line as the first possibility in
the table (with $Z\rightarrow Z_1$) while along the shorter vertical line 
these conditions are given by the fourth possibility there (with the substitutions 
$X_{1,2}\rightarrow Y_{1,2}$, $Y_{1,2}\rightarrow V_{1,2}$ $Z\rightarrow
Z_2$). Imposing the $V_1^\pm=-V_2^\mp$ reflection condition results in
$X_1^\pm=-X_2^\mp$ and for the internal particles the $Q$ values turn out to
be:
\begin{equation}
Q_{Z_1}=2Q+m,\quad Q_{Y_2}=Q_{Y_1}=Q+m,\quad Q_{V_2}=Q_{V_1}=Q,\quad
Q_{Z_2}=m,
\end{equation}
which, for $m\geq 1$, are physically acceptable. Furthermore for the spectral
parameters of the reflecting magnon one finds
\begin{equation}
V_2^++\frac{1}{V_2^+}=-\frac{im}{2g},\qquad 
 V_2^-+\frac{1}{V_2^-}=-\frac{i}{2g}(2Q+m),
\end{equation}
indicating that the internal reflection has indeed a zero at this particular
point. Since naively the diagram would give a second order pole the existence
of this zero reduces it to a first order one. The velocity $v(Y_2)$ is the
same as in the first case of (\ref{ketseb}), while 
\begin{equation}
v(V_2)=-\frac{1}{v(X_2)}=-\frac{N}{i4gL}
\end{equation}
showing that at the two lower vertices the velocity of the internal 
particles points along that of the incoming one thus the diagram is
consistent.  

This way it is demonstrated that (in principle at least) the first order
poles, eq.(\ref{polus}), can be explained in terms of Landau type
diagrams. The systematic analysis - that would require a detailed study of all
possible diagrams and also would involve the the comparison of the residues of
the poles of the Landau diagrams and that of the reflection amplitude 
(\ref{multimagrefl}) - is beyond the scope of the present paper. Similarly the
analysis of the poles of the more complicated amplitudes  describing
(multi)magnon reflections on the $Z=0$ brane \cite{HM} \cite{ABR} is left for
a future investigation.

\section{Boundary state formalism in the magnon problem}

In this section we generalize the boundary state formalism - originally worked
out for relativistic integrable boundary theories by Ghoshal and Zamolodchikov 
\cite{GhosZam} - for the case of magnon reflections.  

\subsection{Summary of boundary state formalism for relativistically invariant 
boundary theories}

Ghoshal and Zamolodchikov showed that the relativistically invariant
integrable boundary theories admit two 
{\sl equivalent} Hamiltonian descriptions. In
the so called \lq\lq open channel'' there is a boundary (represented by the
operator $B$) and the bulk particles (represented by the ZF operators
$A_j^\dagger (\theta )$) 
reflect non trivially on this boundary
\begin{equation}\label{reflmatrix}
A_i^\dagger(\theta )B=R^j_i(\theta )A_j^\dagger (-\theta )B 
\end{equation}
($\theta$ denotes the rapidity of the particle). Exchanging the 
time and space coordinates in the Euclidean version of the theory (i.e. doing
a double Wick rotation) one obtains the so called \lq\lq closed channel''
which is nothing but the periodic bulk model without any boundaries. In this
channel there is a special state, the boundary state $\langle B\vert$, that
carries all the information about the boundary. The requirement that
determines the boundary state is that the correlation functions computed in 
the two channels should be identical:
\begin{equation}
\langle O_{1}(x_{1},y_{1}) ... O_{N}(x_{N},y_{N})\rangle =
{{_{B}\langle 0 \mid {\cal T}_{y}O_{1}(x_{1},y_{1}) ...
O_{N}(x_{N},y_{N})\mid 0 \rangle_{B}} \over {_{B}\langle 0 \mid 0 \rangle_{B}}},\end{equation}
\begin{equation}
\langle O_{1}(x_{1},y_{1}) ... O_{N}(x_{N},y_{N})\rangle =
{\langle B \mid {\cal T}_{x}(O_{1}(x_{1},y_{1}) ... O_{N}(x_{N},y_{N})
\mid 0  \rangle \over \langle B\mid 0 \rangle} 
\end{equation}
The boundary state is a kind of  coherent state that - in the simplest case - 
can be written in terms of the $out$-states as
\begin{equation}\label{bstateout}
\langle B\vert =\langle 0\vert\left( 1+\int\limits_0^\infty d\theta 
K^{ab}(\theta )A_a(-\theta)A_b(\theta )+\dots\right) ,
\end{equation}
where dots stand for the contribution containing higher number of particles. 
 The values of the amplitude describing the 
two particle contribution to the boundary state 
$K^{a b}(\theta)$ at negative real $\theta$  are interpreted as the
coefficients of expansion of $\langle B \mid$ in terms of the $in$-states
$\langle 0\vert A_a(\theta)A_b(-\theta), \ \theta > 0$:
\begin{equation}\label{bstatein}
\langle B\vert =\langle 0\vert\left( 1+\int\limits_0^\infty d\theta 
K^{ab}(-\theta )A_a(\theta)A_b(-\theta )+\dots\right) .
\end{equation}  
Since $A_a(\theta)A_b(-\theta) $ and $A_a(-\theta)A_b(\theta)$ are related by
the $S$ matrix
\begin{equation}
 A_a(\theta)A_b(-\theta) =S_{ab}^{cd}(2\theta )A_d(-\theta)A_c(\theta),
\end{equation}
the consistency of the two ways of writing the first two elements of the
boundary state requires
\begin{equation}
K^{dc}(\theta )=K^{ab}(-\theta ) S_{ab}^{cd}(2\theta ).
\end{equation}           
The importance of this equation is that it provides the boundary crossing
condition (BCC) for the reflection amplitudes once $K^{ab}$ and $R_i^j$ are
related.  This link is provided by the reduction formulae 
from which it follows, that $K^{ab}(\theta )$  
is related to the (analytical continuation of the) reflection amplitude as 
\begin{equation}\label{redukciosof}
K^{ab}(\theta )=R^b_{\bar a}(i\frac{\pi}{2}-\theta ).
\end{equation} 
Furthermore in \cite{GhosZam} it is shown that the boundary Yang-Baxter
equations when combined with (\ref{redukciosof}) and 
the unitarity and crossing properties of the bulk
$S$ matrix are sufficient to guarantee that 
\begin{equation}
[K(\theta ),K(\theta^\prime )]=0,\qquad {\rm where}\qquad K(\theta
)=K^{ab}(\theta )A_a(-\theta)A_b(\theta ),
\end{equation}
and as a consequence the boundary state can be written as
\begin{equation}\label{expform}
\langle B\vert =\langle
0\vert\exp\left(\frac{1}{2}\int\limits_{-\infty}^\infty K(\theta
  )d\theta\right) ,
\end{equation}
without any ordering problems.

\subsection{Boundary state in the magnon problem}

There are two major problems one has to solve when trying to implement the
boundary state formalism in the magnon problem: namely the theory underlying
magnon scattering/reflections is {\sl not relativistically invariant} and also 
the analogues of the reduction formulae are missing. 

As a consequence of the non-relativistic nature of the magnon theory the 
so called \lq\lq mirror'' magnon theory (which is obtained by the double Wick
rotation and is defined in the closed channel) is not equivalent to the 
original (open channel)  one. The mirror magnon theory in the bulk is worked 
out in details in \cite{AFmirror}. There it is shown that the momenta and
energies of the magnon ($p$, $E$) and mirror magnon ($\tilde{p}$, $\tilde{E}$) 
are related through the analytic continuations:
\begin{equation}
p\rightarrow 2i{\rm arcsinh} (\frac{\sqrt{1+\tilde{p}^2}}{4g})=i\tilde{E},\qquad 
E=\sqrt{1+16g^2\sin^2\frac{p}{2}}\rightarrow i\tilde{p} .
\end{equation}
Realizing that the dispersion relation (\ref{disp}) describes a complex torus 
\cite{BHL,Janik:2006dc} the magnon energy, momentum or equivalently the
spectral parameters $x^\pm$ can be expressed in terms of Jacobi elliptic
functions:
\begin{equation}
p(z)=2{\rm am}(z),\qquad \sin\frac{p(z)}{2}={\rm sn}(z,k)\equiv {\rm sn}(z),
\qquad 
E(z)={\rm dn}(z,k)\equiv{\rm dn}(z),
\end{equation}
\begin{equation}
x^\pm(z)=\frac{1}{4g}\left(\frac{{\rm cn}(z,k)}{{\rm sn}(z,k)}\pm i\right)
\left(1+{\rm dn}(z,k)\right) .
\end{equation}
Here the elliptic modulus $k^2=-16g^2\in {\mathbb R}$ is fixed in a given
theory thus $p$, $E$ or $x^\pm$ can be regarded as functions of the complex 
parameter $z$ called \lq\lq generalized rapidity''. The two cycles of the
rapidity torus can be described by the shifts $z\rightarrow z\pm 2\omega_1$,  
$z\rightarrow z\pm 2\omega_2$ with 
\begin{equation}
\omega_1=2K(k^2),\qquad \omega_2=2iK(1-k^2)-2K(k^2),
\end{equation}
where $K(k^2)$ is the complete elliptic integral of the first kind. For our 
$k^2$-s ${\rm Im}\,\omega_1=0={\rm Re}\,\omega_2$. To make contact with
relativistic theories we consider  the limit
$g\rightarrow\infty$ when the periods of the torus have the following behaviour
\begin{equation}
\omega_1\rightarrow\frac{\log g}{2g},\qquad \omega_2\rightarrow
i\frac{\pi}{4g}.
\end{equation}
Rescaling $z$ as $z\rightarrow z/(4g)$ and the momentum as $p\rightarrow
p/(2g)$ keeps the range of Im$(z)$ finite and converts also the the dispersion
relation (\ref{disp}) to the relativistic form $E^2-p^2=1$ thus showing that 
the variable $z$ indeed plays the role of $\theta$ (because $p=\sinh z$).  
Furthermore in this limit the torus degenerates into the strip $-\pi <{\rm
  Im}(z) <\pi$ and $-\infty <{\rm Re}(z)<\infty$. In \cite{AFmirror} it is
shown that the magnon 
$S$-matrix (\ref{ssmat}) admits an analytic continuation ${\cal S}(z_1,z_2)$ 
to the entire rapidity torus.

The $z$-torus can also be used to describe the mirror model. The trick is to 
realize that the double Wick rotation can be implemented \cite{AFmirror} 
by the shift  $z\rightarrow \tilde{z}+\frac{\omega_2}{2}$ since then 
\begin{equation}
\tilde{p}=-i{\rm dn}(\tilde{z}+\frac{\omega_2}{2},k)=\sqrt{k^\prime}
\frac{{\rm sn}(\tilde{z})}{{\rm cn}(\tilde{z})}, \qquad
\tilde{E}=2{\rm arccoth}\frac{\sqrt{k^\prime}}{{\rm dn}(\tilde{z})},
\end{equation}
i.e. $\tilde{p}$ is real for real $\tilde{z}$. Furthermore the range $-\infty 
<\tilde{p} <\infty$ corresponds to $-\omega_1/2<\tilde{z} <\omega_1/2$. 
Note also that this shift is completely analogous to
$\theta\rightarrow\theta+i\frac{\pi}{2}$ connecting the rapidities of
particles in the open and closed channels in case of relativistic theories.  
According to Arutyunov and Frolov the mirror magnon's scattering matrix 
$\tilde{\cal S}(\tilde{z}_1,\tilde{z}_2)$ is related to the magnon $S$ matrix
as \cite{AFmirror}:
\begin{equation}\label{mirrorS}
 \tilde{\cal S}(\tilde{z}_1,\tilde{z}_2)={\cal S}(\tilde{z}_1+
\frac{\omega_2}{2},\tilde{z}_2+\frac{\omega_2}{2}).
\end{equation}

\subsubsection{Boundary state for magnon reflections on the $Y=0$ brane}

The simple description of boundary state given by (\ref{bstateout},\ref{bstatein})
can immediately be generalized to magnons reflecting on the $Y=0$ brane since 
this problem  
contains no boundary degrees of freedom (and no boundary bound
states). Restricting our attention to magnon reflection matrices corresponding
to a single copy of the centrally extended $su(2\vert 2)$
algebra\footnote{with the understanding that the complete reflection matrix is
  the tensor product of two such $R$-s}  
the analogue
of (\ref{reflmatrix}) is 
\begin{equation}\label{magreflmatrix}
A_i^\dagger(z )B=R^j_i(z )A_j^\dagger (-z )B 
\end{equation} 
where the indices $i,j=1\dots 4$. Symmetry considerations restrict the
explicit form of the reflection matrix as $R^j_i(z )=R_0(z)\,{\rm
  diag}(e^{-ip(z)},-1,1,1)$ \cite{HM} \cite{AN1}.  
Denoting the ZF operators for the mirror
magnons as $\tilde{A}_a(\tilde{z})$ the boundary state can be written in terms
of the generalized (shifted) rapidity as 
\begin{equation}\label{magbstateout}
\langle B\vert =\langle 0\vert\left( 1+\int\limits_0^{\omega_1/2} d\tilde{z} 
\rho(\tilde{z}) K^{ab}(\tilde{z} )\tilde{A}_a(-\tilde{z})
\tilde{A}_b(\tilde{z})+\dots\right)
\end{equation}
where $\rho(\tilde{z})$ is the density of states that plays no role in our
considerations. The consistency condition of the two ways of expressing the
boundary state has the form now:
\begin{equation}\label{magconsistency}
K^{dc}(\tilde{z})=K^{ab}(-\tilde{z}) \tilde{S}_{ab}^{cd}(\tilde{z},-\tilde{z}),
\end{equation}    
i.e. naturally it contains the $S$ matrix of the mirror model.

 (\ref{magconsistency}) can be interpreted 
 as the BCC for the reflection matrix if 
$K^{ab}$ and $R_i^j$ are somehow related. In the lack of  reduction formulae in
the magnon model we determine a relation between them by demanding that the 
$z\rightarrow z+\frac{\omega_2}{2}$ continuation of the boundary Yang-Baxter 
equation for $R_i^j$ 
\begin{equation}
R_j^k(z_2)S_{ik}^{lm}(z_1,-z_2)R_l^n(z_1)S_{mn}^{wv}(-z_2,-z_1)=
S_{ij}^{mn}(z_1,z_2)R_m^q(z_1)S_{nq}^{uv}(z_2,-z_1)R_u^w(z_2) ,
\end{equation}
when combined with the unitarity and crossing properties of the
magnon $S$ matrix 
\begin{equation}\label{bulkstulajd}
{\cal S}_{12}(z_1,z_2){\cal S}_{21}(z_2,z_1)={\mathbb I},\qquad
{\mathbb C}^{-1}{\cal S}_{12}^{tr}(z_1,z_2){\mathbb C}{\cal
  S}_{12}(z_1,z_2-\omega_2)={\mathbb I},
\end{equation}
and with the AF relation (\ref{mirrorS}) between the magnon
and mirror magnon's $S$ matrices should guarantee that
\begin{equation}\label{magkom}
[K(z),K(z^\prime )]=0,\qquad {\rm where}\qquad K(z
)=K^{ab}(z )\tilde{A}_a(-z)\tilde{A}_b(z ).
\end{equation}
After a not very illuminating computation one finds that this condition is met
if
\begin{equation}\label{magKR}
K^{ab}(z)={\mathbb C}^{ac}R_c^b(\frac{\omega_2}{2}-z)
\end{equation}     
where ${\mathbb C}$ is the charge conjugation matrix. Note the complete
analogy of this equation to (\ref{redukciosof}) obtained by using the
reduction formulae.

Since (\ref{magKR}) is obtained by requiring (\ref{magkom}) the boundary state
has a similar exponential form as in the BIQFT case
(\ref{expform}). Furthermore,  
plugging (\ref{magKR})
into (\ref{magconsistency}), using eq.(\ref{mirrorS}) and continuing back by
substituting $z\rightarrow u+\frac{\omega_2}{2}$ gives
\begin{equation}
{\mathbb C}^{ac_1}R_{c_1}^b({\omega_2}+u)S_{ab}^{cd}(u+\omega_2,-u)=
{\mathbb C}^{dm}R_m^c(-u).
\end{equation}
Using the unitarity of the reflection matrix $R(u)R(-u)={\mathrm 1}$ and the
$S(z_1+\omega_2 ,z_2+\omega_2)=S(z_1,z_2)$ \lq\lq translational invariance''
property of the magnon $S$ matrix  
this last
equation can be converted to
\begin{equation}
{\mathbb C}^{ac_1}R_{c_1}^b({\omega_2}+u)S_{ab}^{cd}(u,-u-\omega_2)R_c^n(u)=
{\mathbb C}^{dn},
\end{equation}
which, in the light of $x^\pm(u+\omega_2)=(x^\pm(u))^{-1}$,  
is the BCC in \cite{AN1}.  

\subsubsection{Boundary state for magnon reflections on the $Z=0$ brane}

The essentially new feature of the magnon reflection on the $Z=0$ brane is the
presence of boundary degrees of freedom \cite{HM}. This means that 
in this case the
boundary is not a singlet but belongs to the same fundamental representation 
of the symmetry algebra as the magnons themselves (albeit with a slightly 
different relation between the central charges). As a consequence in a
reflection process also the boundary may change and eq.(\ref{magreflmatrix}) 
generalizes to 
\begin{equation}
 A_i^\dagger(z )B_\alpha =R^{j\beta }_{i\alpha }(z )A_j^\dagger (-z )B_\beta
\end{equation}
where $B_\alpha $ $\alpha =1\dots 4$ are the operators representing the $Z=0$ 
boundary. The explicit form of the reflection matrix as obtained from imposing 
the symmetry requirement is given in eq.(3.32) and (3.33) of \cite{AN1}.

Looking at the graphical representation of the magnon reflection together with
its image in the closed channel 

\vspace{.5cm}~~~~~~~~~~~~~~~~~~~~~~~~~~~~~~~~~~~~~~\includegraphics[%
  width=8.5cm,
  height=4.5cm]{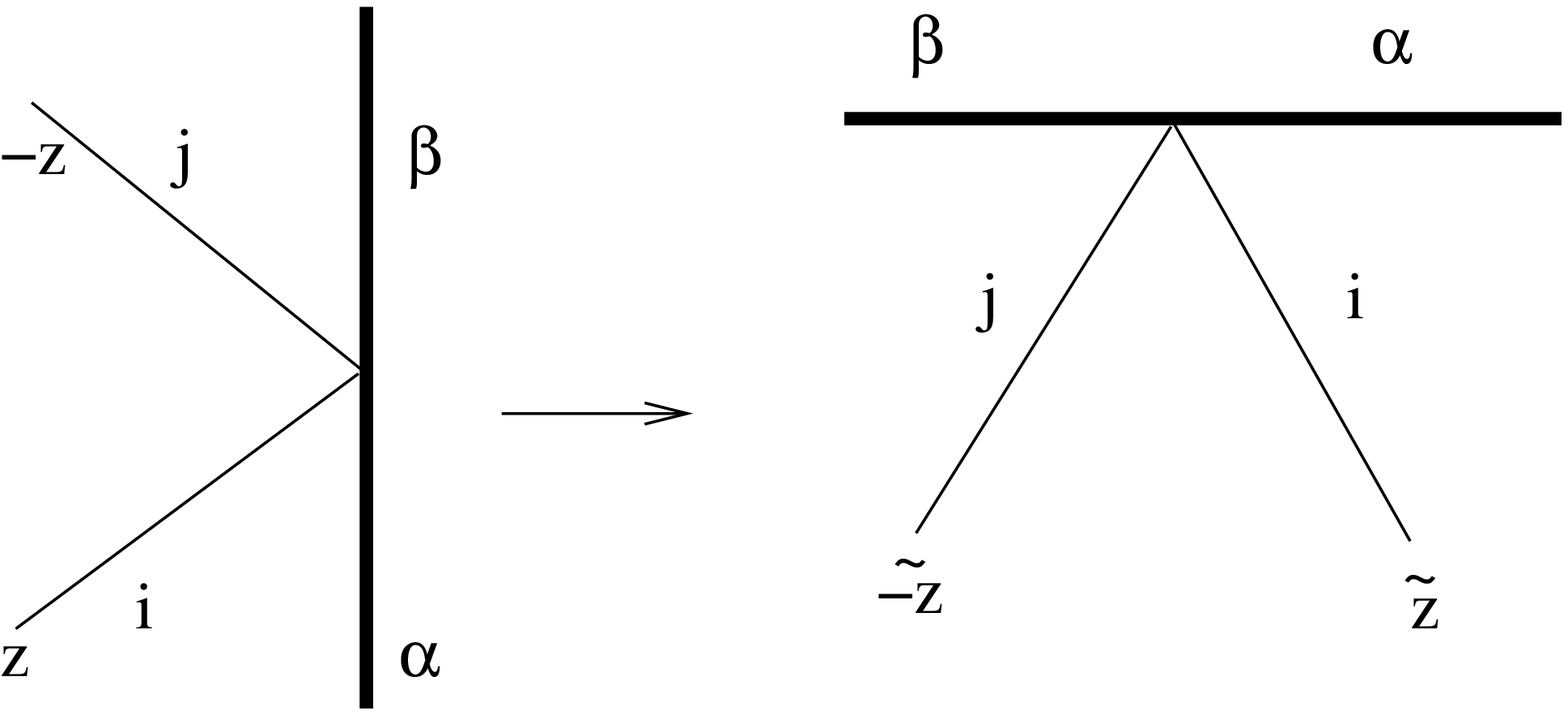}

\noindent it is easy to argue that in the closed channel one indeed obtains the
periodic mirror magnon model but with several 
sectors characterized by the boundary
degrees of freedom $\alpha ,\beta $. In each sector there is a boundary state
\begin{equation}\label{magbstatez}
\langle B_{\alpha\beta}\vert =\langle 0\vert ( \delta_{\alpha\beta}+
\int\limits_0^{\omega_1/2} d\tilde{z} 
\rho(\tilde{z})K^{\alpha\beta}(\tilde{z}) +
\int\limits_0^{\omega_1/2} d\tilde{z}\rho(\tilde{z})
\int\limits_0^{\omega_1/2} d\tilde{w}\rho(\tilde{w})
K^{\alpha\gamma}(\tilde{z})K^{\gamma\beta}(\tilde{w})
+\dots ) 
\end{equation}
where
\begin{equation}\label{magbstatez2}
K^{\alpha\beta}(\tilde{z})=
K^{ab\alpha\beta }(\tilde{z} )\tilde{A}_a(-\tilde{z})
\tilde{A}_b(\tilde{z}) .
\end{equation}
The consistency condition of the two ways of expressing the
boundary state has the form now:
\begin{equation}\label{magconsistencyz}
K^{dc\alpha\beta }(\tilde{z})=K^{ab\alpha\beta }(-\tilde{z})
\tilde{S}_{ab}^{cd}(\tilde{z},-\tilde{z})
\end{equation}
Note that both the boundary state (\ref{magbstatez}) 
and the consistency condition (\ref{magconsistencyz}) are simply obtained by 
\lq\lq decorating'' with the indeces $\alpha\  \beta$ the corresponding
expressions for the $Y=0$ case, eq.(\ref{magbstateout} \ref{magconsistency}).

Recalling that the analytically continued boundary Yang Baxter equation - that
graphically can be represented as - 

\vspace{.5cm}~~~~~~~~~~~~~~~~~~\includegraphics[%
  width=11.5cm,
  height=2.5cm]{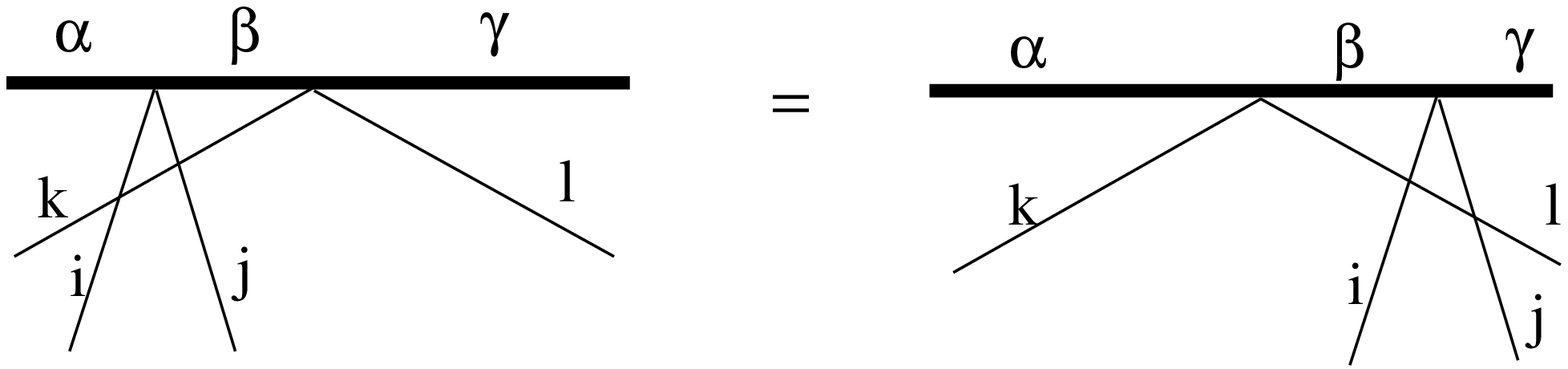}

\noindent contains a summation over the intermediate boundary degree of
freedom $\beta $ makes it plausible that demanding the vanishing of
\begin{equation}
[K(\tilde{z}),K(\tilde{w})]^{\alpha\beta}\equiv K^{\alpha\gamma}(\tilde{z})
K^{\gamma\beta}(\tilde{w})-K^{\alpha\gamma}(\tilde{w})
K^{\gamma\beta}(\tilde{z})=0
\end{equation}
when combined with eq.(\ref{mirrorS}) and eq.(\ref{bulkstulajd}) gives indeed 
a useful relation between the reflection matrix and the coefficient of the two 
particle contribution to the boundary state. This way one obtains
\begin{equation}
K^{ab\alpha\beta }(z)={\mathbb C}^{ac}R_{c\alpha }^{b\beta
}(\frac{\omega_2}{2}-z)
\end{equation}
in complete analogy to eq.(\ref{magKR}) and (\ref{redukciosof}). Proceeding in
the same way as in the case of reflections on the $Y=0$ brane -
i.e. continuing back by the substitution $z\rightarrow u+\frac{\omega_2}{2}$ 
and exploiting the unitarity of $R^{a\alpha}_{b\beta}$ and the translational 
symmetry of the magnon scattering matrix - one can show that the consistency
condition, eq.(\ref{magconsistencyz}), becomes indeed the BCC for the $Z=0$
brane, eq.(4.24) of \cite{AN1}. 
 
\section{Conclusions}

In this paper two problems related to reflections of (multi)magnons in AdS/CFT 
are discussed. In the first problem, aimed at giving an interpretation in
terms of Landau equations and Landau diagrams of the poles of the reflection
amplitudes that do not correspond to boundary bound states we pointed out that
the derivation of Landau equations for the magnon problem requires the
knowledge of the free propagator for the (multi)magnons. Using an appropriate
candidate for this propagator the Landau equations were derived and some
differences to the ordinary case were pointed out. As a result of these
differences the singularities of the the magnon reflection/scattering
amplitudes may be interpreted in terms of space time (Landau) diagrams, but - 
unlike in the relativistic case - these diagrams do not correspond to the 
propagation of on shell particles. In addition a detailed study of Landau
diagrams describing the first order poles of the magnon reflection amplitudes
on the $Y=0$ brane is presented. 

The boundary state formalism originally worked out for relativistic boundary
integrable models by Ghoshal and Zamolodchikov is succesfully generalized to 
magnon reflections on both the $Y=0$ and the $Z=0$ branes. This way a new
derivation is obtained of the boundary crossing condition and this is
interesting, as this condition plays an important role in determining the
scalar factor of the reflection amplitudes which is left undetermined by the
symmetry considerations. In addition 
the boundary states constructed may be useful 
to investigate the finite size effects (TBA) of magnon reflections.
 
\subsection*{Acknowledgments}

The author thanks Dr. Z. Bajnok for the numerous 
illuminating discussions and the participants of the Focus program on 
\lq\lq Finite size technology in low dimensional quantum systems IV'' (24 June
- 15 July 2008, APCTP, Pohang Korea)
for the interesting remarks. 
This research was partially supported by the Hungarian
research fund OTKA K60040.

\end{document}